\documentclass[conference]{IEEEtran}
\IEEEoverridecommandlockouts

\usepackage{cite}
\usepackage{amsmath,amssymb,amsfonts}
\usepackage{algorithmic}
\usepackage{graphicx}
\usepackage{textcomp}
\usepackage{xcolor}
\usepackage{multirow}
\usepackage{hyperref}
\usepackage{float}
\usepackage{booktabs}
\usepackage{subcaption}

\def\BibTeX{{\rm B\kern-.05em{\sc i\kern-.025em b}\kern-.08em
    T\kern-.1667em\lower.7ex\hbox{E}\kern-.125emX}}
\begin{document}

\title{
Retrieval-Augmented Classification of Environmental Mitigations in Hydropower Licensing Documents
\thanks{
This manuscript has been authored by UT-Battelle, LLC under Contract No. DE-AC05-00OR22725 with the U.S. Department of Energy. The United States Government retains and the publisher, by accepting the article for publication, acknowledges that the United States Government retains a non-exclusive, paid-up, irrevocable, world-wide license to publish or reproduce the published form of the manuscript, or allow others to do so, for United States Government purposes. The Department of Energy will provide public access to these results of federally sponsored research in accordance with the DOE Public Access Plan (http://energy.gov/downloads/doe-public-access-plan).
}
}

\author{
\IEEEauthorblockN{
Hong-Jun Yoon\IEEEauthorrefmark{1},
Tom Ruggles\IEEEauthorrefmark{2},
Joanna Lee\IEEEauthorrefmark{3} and
Debjani Singh\IEEEauthorrefmark{2}
}

\IEEEauthorblockA{
\IEEEauthorrefmark{1}Computational Sciences and Engineering Division, 
Oak Ridge National Laboratory, Oak Ridge, Tennessee 37830, USA \\
}
\IEEEauthorblockA{
\IEEEauthorrefmark{2}Environmental Sciences Division, 
Oak Ridge National Laboratory, Oak Ridge, Tennessee 37830, USA \\
}
\IEEEauthorblockA{
\IEEEauthorrefmark{3}Farragut High School, Knoxville, Tennessee 37934\\
}

}

\maketitle

\begin{abstract}
Identifying and classifying environmental mitigation obligations in Federal Energy Regulatory Commission hydropower licensing documents is a labor-intensive task requiring deep domain expertise. We formulate this as a multi-label classification problem over a structured 135-category taxonomy and address the central challenge of severe label scarcity: 40 of 135 categories have no training examples, and 26 have fewer than five. A supervised Bidirectional Encoder Representations from Transformers (BERT)-based pipeline, while effective on well-represented categories, achieves F1 of zero on unseen classes regardless of augmentation strategy. We introduce a Retrieval-Augmented Generation (RAG) pipeline that conditions classification on retrieved category definitions, enabling zero-shot generalization across the full label space. We further propose a hybrid system that combines BERT detection with RAG classification, exploiting the high recall of fine-tuned detection and the zero-shot coverage of retrieval-augmented reasoning. Evaluated on the full set of 2017 license documents (5,860 paragraphs, 135 categories), the hybrid achieves a Micro F1 of 0.524, outperforming the BERT-only pipeline (0.477) and the RAG-only pipeline (0.416) across all training-support buckets.
\end{abstract}


\section{Introduction}
\label{sec:intro}
The Federal Energy Regulatory Commission (FERC) is the primary U.S. federal agency responsible for licensing non-federal hydropower projects, overseeing more than 2,000 facilities nationwide. As a condition of each license, FERC mandates a broad set of environmental mitigation measures — ranging from fish passage infrastructure and water quality monitoring to habitat restoration and species protection plans. These requirements are codified as mitigation obligations scattered throughout the license documents themselves, documents that routinely exceed 15,000 words per project.

To support regulatory oversight and environmental research, ORNL has developed a structured taxonomy of 135 mitigation categories, organized under six resource groups including fish passage, water quality, and biodiversity\cite{schramm2016synthesis,ruggles2025comprehensive}. Each paragraph of a FERC license document may reference zero, one, or multiple categories from this taxonomy, making the extraction task a fine-grained multi-label classification problem over dense, domain-specific text. Manually compiling this information is both labor-intensive and error-prone: trained annotators spend over three hours per document, and the subjective nature of category boundaries introduces considerable inter-reader variability.

The central challenge in automating this process is severe label imbalance. Of the 135 mitigation categories, 40 have no training examples whatsoever and 26 have fewer than five — a long-tail distribution that reflects the genuine rarity of certain mitigation types across the licensed project corpus. In our prior work \cite{yoon2024bigdata}, we demonstrated that a BERT-based detection and classification pipeline performs reasonably well on frequently observed categories. To address the long tail, we further explored augmenting the training set with mitigation category descriptions and synthesizing additional examples for underrepresented classes. While these efforts yielded marginal improvements, they did not constitute a fundamental solution: a supervised model cannot predict a category it has never seen, and no augmentation strategy fully compensates for the absence of real annotated data.

Retrieval-Augmented Generation (RAG) offers a principled path forward. Rather than relying on learned associations from scarce training data, a RAG system retrieves relevant category definitions from the structured taxonomy and presents them as context to a large language model (LLM), which then reasons over the paragraph to determine applicability. This approach requires no category-specific training examples and naturally extends to all 135 categories, including those absent from the training set. We describe the RAG framework in detail in Section II.

Building on this insight, we propose a hybrid pipeline that pairs BERT detection — which achieves high recall across common categories — with RAG classification, which provides zero-shot coverage for the long tail. Evaluated on the full set of 2017 FERC license documents (5,860 paragraphs, 135 categories), the hybrid achieves a Micro F1 of 0.524, outperforming both the BERT-only pipeline (0.477) and the RAG-only pipeline (0.416) across every training-support bucket.

The remainder of this paper is organized as follows. Section II provides background on RAG and its application to multi-label document classification. Section III describes the dataset, models, implementation, and evaluation metrics. Section IV presents experimental results. Section V discusses findings and limitations, and Section VI concludes. Our main contributions are:

\begin{itemize}
    \item [$\bullet$] A RAG-based classification pipeline that generalizes to zero-shot and rare mitigation categories unreachable by supervised methods alone.
    \item [$\bullet$] A hybrid system (BERT detection + RAG classification) that outperforms either component across all training-support buckets, with particular strength on the long tail of underrepresented categories.
\end{itemize}

\section{Background}
\label{sec:background}
Retrieval-Augmented Generation (RAG) was introduced by Lewis et al. \cite{lewis2020rag} as a framework for knowledge-intensive NLP tasks in which the information required to process a query cannot be fully encoded in model parameters alone. The core idea is to augment a generative language model with a non-parametric retrieval component: given an input query, a retriever identifies relevant passages from an external knowledge base, and a large language model (LLM) conditions its output on both the query and the retrieved context. This decoupling of knowledge storage from model parameters allows the system to leverage domain-specific information without retraining, and has been shown to substantially reduce hallucination in knowledge-intensive settings \cite{gao2024ragsurvey}.

While RAG was originally applied to open-domain question answering, subsequent work has demonstrated its effectiveness for text classification tasks. Chalkidis and Kementchedjhieva \cite{chalkidis2023retrieval} applied retrieval augmentation to multi-label text classification in the legal and biomedical domains, showing that augmenting document representations with similar training instances substantially improves performance on infrequent labels. More recently, \cite{ictir2024zeroshot} proposed a training-free RAG approach for zero-shot text classification, enriching query representations with retrieved category knowledge to extend classification to unseen classes without any model retraining. These works establish that retrieval augmentation is particularly well-suited to the long-tail regime, where supervised methods degrade due to label scarcity.

Multi-label classification of regulatory and legislative documents has been studied extensively in the legal NLP community. Chalkidis et al. \cite{chalkidis2019eurlex} introduced the EURLEX57K benchmark of 57,000 EU legislative documents tagged with approximately 4,300 EUROVOC concepts, establishing that label imbalance and zero-shot categories are endemic to legal document classification — a finding that directly motivates our work. Domain-adapted language models such as LEGAL-BERT \cite{chalkidis2020legalbert}, pre-trained on large corpora of legal text, have been shown to outperform general-purpose BERT on downstream legal classification tasks. Our work uses the general \texttt{bert-base-uncased} model, as the FERC licensing domain is narrower and more specialized than broad legal corpora; the impact of domain-specific pre-training on this task remains an open question.

The challenge of long-tail label distributions is not unique to the legal domain. In patent classification — another fine-grained, multi-label task with severe class imbalance — recent work has demonstrated that LLMs under retrieval-augmented prompting outperform fine-tuned encoder models on rare and zero-shot categories, while encoder models retain their advantage on frequent ones \cite{patent2026longtail}. This complementarity motivates the hybrid design we adopt in this paper.

In the regulatory NLP domain, prior work on FERC hydropower documents focused on rule-based extraction and expert-driven cataloguing of mitigation requirements \cite{schramm2016synthesis}. Our previous system \cite{yoon2024bigdata} introduced a BERT-based detection and classification pipeline for this task, demonstrating the feasibility of automated mitigation extraction and establishing a baseline over the 135-class taxonomy. The present work extends this line of research by incorporating RAG to address the zero-shot and rare-category limitations of the supervised approach.

\section{Methods}
\label{sec:methods}
We evaluate three pipeline configurations on a shared dataset of FERC hydropower licensing documents: a BERT-based pipeline, a RAG pipeline, and a hybrid that combines the two. The following subsections describe the dataset, the models, their implementation, and the evaluation metrics.

\subsection{Data}
Our dataset consists of FERC hydropower licensing documents spanning the years 2014 through 2017, collected and annotated by domain experts at ORNL. Each document is segmented at the paragraph level, and each paragraph is annotated with zero or more mitigation category IDs drawn from a structured 135-class taxonomy organized under six resource groups: fish passage, water quality, biodiversity, habitat, recreation, and administrative. The annotation task is therefore a multi-label classification problem; 19.4\% of annotated positive paragraphs carry more than one category label.

Documents from 2014 through 2016 are used for training and validation (90/10 split), and documents from 2017 constitute the held-out test set. The 2017 test set comprises 5,860 paragraphs, of which 753 are annotated positive (12.8\%), across 21 license documents. This represents an expansion from our prior work \cite{yoon2024bigdata}, which evaluated on a subset of 2017 documents totaling 3,247 paragraphs; the additional documents bring greater category diversity and a more representative evaluation benchmark.

The label distribution is severely skewed. Table~\ref{tab:label_dist} summarizes the distribution of training support across the 135 categories. Forty categories have no training examples at all, and 26 have fewer than five — together accounting for nearly half of the taxonomy. This long-tail structure is the central challenge the present work addresses.

\begin{table}[h]
\centering
\caption{Training support distribution across 135 mitigation categories.}
\label{tab:label_dist}
\begin{tabular}{lcc}
\hline
Support Bucket & \# Categories & \# in Test \\
\hline
Zero-shot (0 examples)  & 41 & 2  \\
Rare (1--5 examples)    & 26 & 6  \\
Medium (6--20 examples) & 34 & 12 \\
Common ($>$20 examples) & 34 & 33 \\
\hline
Total                   & 135 & 53 \\
\hline
\end{tabular}
\end{table}

Notably, only 53 of the 135 categories appear in the 2017 test set. The implications of this coverage gap for metric selection are discussed in Section~\ref{sec:metrics}.

\subsection{Models}
\begin{figure*}[t]
    \centering
    \includegraphics[width=0.85\textwidth]{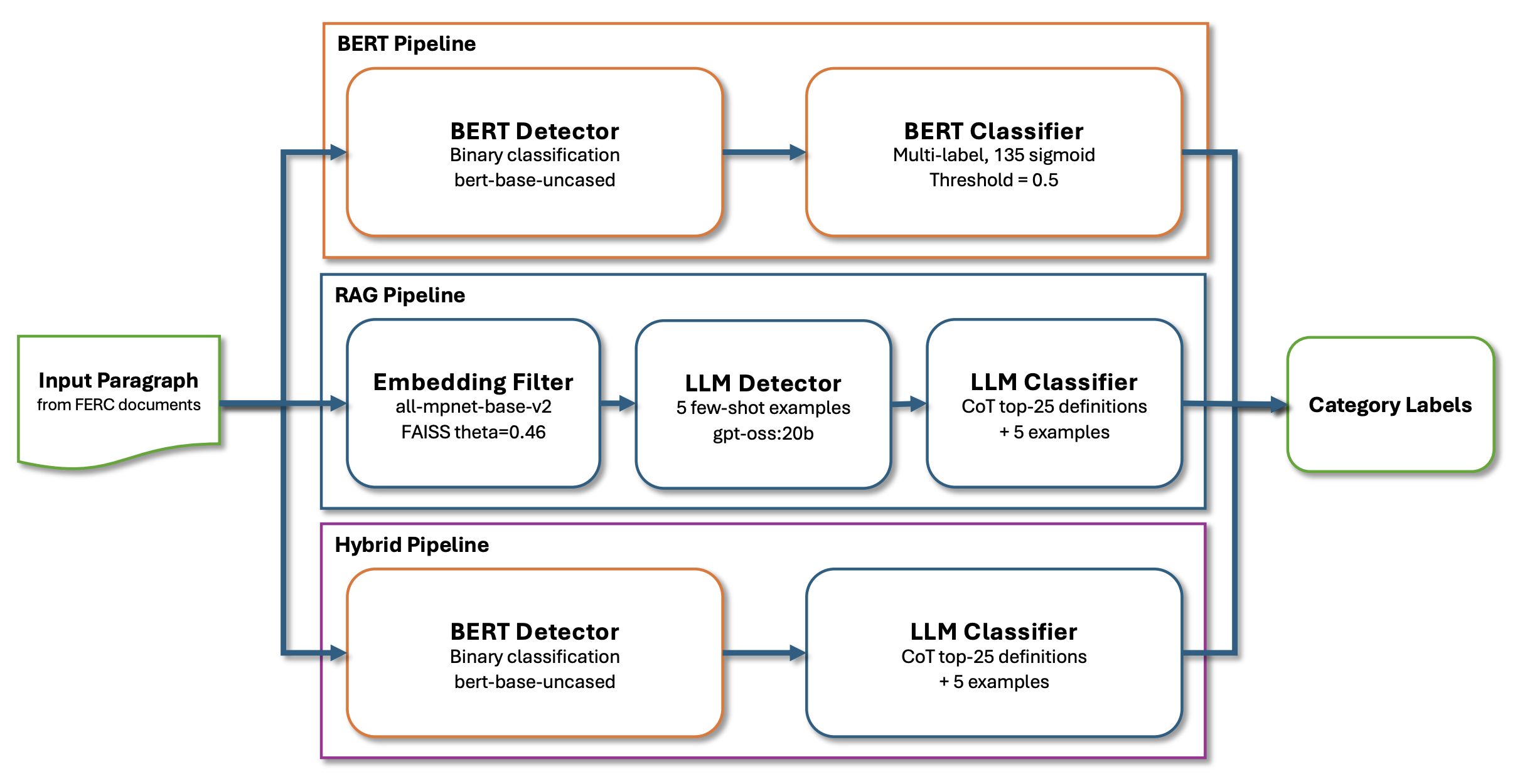}
    \caption{Architecture of the three evaluated pipelines. The BERT pipeline
    (top) applies a fine-tuned binary detector followed by a multi-label
    classifier. The RAG pipeline (middle) uses embedding-based filtering and
    LLM inference conditioned on retrieved category definitions. The hybrid
    pipeline (bottom) combines BERT detection with RAG classification,
    bypassing the LLM detection step.}
    \label{fig:architecture}
\end{figure*}

All three systems share a two-stage architecture consisting of a detection stage followed by a classification stage. The separation of these two stages reflects a fundamental difference in what each task is asking. Detection addresses whether a paragraph contains a forward-looking environmental mitigation obligation — that is, whether the license is mandating that a specific environmental condition be achieved or maintained. Classification, applied only to detected paragraphs, addresses which mitigation category or categories that obligation belongs to: fish passage, water quality monitoring, habitat restoration, and so forth. Because these are semantically distinct questions requiring different linguistic and domain cues, they are modeled separately \cite{yoon2024bigdata}. As a practical consequence, this design also avoids applying a 135-class classifier to the large majority of paragraphs that carry no mitigation content. The architecture of all three pipelines is illustrated in Fig.~\ref{fig:architecture}.

\subsubsection{BERT Pipeline}
The BERT pipeline follows the architecture introduced in \cite{yoon2024bigdata}. The detection model is a binary sequence classifier built on \texttt{bert-base-uncased} \cite{devlin2019bert}, fine-tuned to predict whether a paragraph contains any mitigation obligation. The classification model is a multi-label classifier, also based on \texttt{bert-base-uncased}, with 135 sigmoid output nodes over the full category set, allowing a paragraph to be assigned any combination of category labels simultaneously. A category is predicted as present if its sigmoid output exceeds a threshold of 0.5. Both models are trained on the 2014--2016 documents and evaluated end-to-end on the 2017 test set.

\subsubsection{RAG Pipeline}
The RAG pipeline replaces both BERT stages with retrieval-augmented LLM inference. Given a paragraph, a sentence transformer model computes a dense embedding that is compared against precomputed embeddings of all 135 category definitions stored in a FAISS index \cite{johnson2019faiss}. The top-$k$ most semantically similar category definitions are retrieved and passed as context to a large language model, which performs chain-of-thought reasoning to determine first whether the paragraph contains a mitigation obligation and then which of the retrieved categories apply. Five annotated examples are included in the prompt to provide task grounding. Because classification is conditioned on retrieved category definitions rather than learned label embeddings, the RAG pipeline generalizes naturally to zero-shot and rare categories.

\subsubsection{Hybrid Pipeline}
The hybrid pipeline combines the complementary strengths of both approaches. BERT detection serves as a recall-oriented filter, efficiently identifying candidate paragraphs from the full document. RAG classification then operates on those candidates, leveraging retrieved category definitions to handle the full label space including zero-shot and rare categories. Paragraphs flagged as positive by the BERT detector are passed directly to the RAG classifier, bypassing the LLM detection step. This design preserves the computational efficiency of learned binary filtering while delegating the harder categorization problem to retrieval-augmented reasoning.

\subsection{Experimental Setup}

BERT models: Both the detector and the classifier are initialized from \texttt{bert-base-uncased} \cite{devlin2019bert} and fine-tuned on the 2014--2016 training documents. Training uses the AdamW optimizer with a learning rate of $2 \times 10^{-5}$, weight decay of 0.01, and a batch size of 16, with a linear warmup over the first 10\% of training steps. Models are trained for up to 50 epochs with early stopping based on validation performance. The classification threshold for the multi-label classifier is fixed at 0.5.

RAG pipeline: Paragraph and category definition embeddings are computed using \texttt{all-mpnet-base-v2} from the Sentence Transformers library \cite{reimers2019sentencebert}, and stored in a FAISS flat inner-product index \cite{johnson2019faiss}. At inference time, the top $k=25$ category definitions are retrieved for each paragraph. Paragraphs whose maximum embedding similarity to any category definition falls below a threshold of 0.46 are discarded without LLM inference, serving as a fast pre-filter. For paragraphs that pass this filter, the retrieved definitions along with five annotated few-shot examples are passed to a 20-billion parameter open-weight LLM served locally via Ollama. The same model handles both the detection and classification steps to avoid GPU memory swapping between stages.

\subsection{Performance Metrics}
\label{sec:metrics}
We report both Micro F1 and Macro F1 for the end-to-end pipeline evaluation, but treat Micro F1 as the primary metric. Micro F1 aggregates true positives, false positives, and false negatives at the instance level across all categories, and thus reflects the system's practical utility without penalizing for categories that are structurally absent from the evaluation set. Macro F1, by contrast, computes F1 independently for each of the 135 categories and averages the results equally. Because only 53 of the 135 categories appear in the 2017 test set, categories with no test instances contribute an F1 of zero by definition, imposing a theoretical ceiling of $53/135 \approx 0.393$ on Macro F1 regardless of model quality. This ceiling is a property of the data split rather than a model limitation and must be interpreted accordingly.

For the BERT detector, we report precision, recall, and F1 separately, as the trade-off between these quantities has direct implications for downstream classification performance.

To understand how each system handles the label scarcity problem, we additionally report Macro F1 stratified by training support bucket — zero-shot, rare, medium, and common — computed only over categories that appear in the test set. This analysis isolates the contribution of each system to the long-tail coverage problem identified in Section~\ref{sec:intro}.

\section{Results}
\label{sec:results}
We evaluate all three pipelines on the 2017 test set and report results in terms of Micro F1 (primary), Macro F1, and per-bucket Macro F1 over categories present in the test set.

\subsection{Overall Performance}
Table~\ref{tab:main_results} compares the four system configurations. The hybrid pipeline achieves the highest Micro F1 of 0.524, outperforming the BERT-only pipeline (0.477) and both RAG configurations. Among the RAG variants, providing five few-shot examples yields a consistent improvement over zero-shot prompting (0.416 vs. 0.350), confirming the value of in-context grounding. All Macro F1 scores remain well below the theoretical ceiling of 0.396, reflecting the difficulty of the task under sparse label coverage.

\begin{table}[h]
\centering
\caption{End-to-end pipeline performance on the 2017 test set.}
\label{tab:main_results}
\begin{tabular}{lcc}
\hline
System & Micro F1 & Macro F1 \\
\hline
RAG, zero-shot              & 0.350 & 0.121 \\
RAG, 5 few-shot examples    & 0.416 & 0.141 \\
BERT pipeline               & 0.477 & 0.129 \\
Hybrid (BERT + RAG)         & \textbf{0.524} & \textbf{0.160} \\
\hline
\end{tabular}
\end{table}

\subsection{BERT Detector Performance}
Table~\ref{tab:detector} reports the performance of the BERT detector used in both the BERT pipeline and the hybrid. The detector achieves high recall (0.920), missing only 60 of 753 positive paragraphs, at the cost of 406 false positives passed to the downstream classifier. This asymmetry is intentional: in the context of a licensing compliance task, missing a mitigation obligation is more costly than over-generating candidates for classification.

\begin{table}[h]
\centering
\caption{BERT detector performance on the 2017 test set.}
\label{tab:detector}
\begin{tabular}{cccccc}
\hline
Pr & Re & F1 & TP & FP & FN \\
\hline
0.631 & 0.920 & 0.748 & 693 & 406 & 60 \\
\hline
\end{tabular}
\end{table}

\subsection{Performance by Training Support}
Table~\ref{tab:bucket} reports Macro F1 stratified by training support bucket, computed only over categories present in the 2017 test set. The hybrid outperforms both components in every bucket. BERT achieves F1 of zero on zero-shot and rare categories by construction — it cannot predict labels it has never seen. RAG covers these categories through definition-driven reasoning, and the hybrid amplifies this advantage by reducing the false positive load passed to the classifier. On common categories ($>$20 training examples), BERT and the hybrid perform comparably, with RAG slightly behind.

\begin{table}[h]
\centering
\caption{Macro F1 by training support bucket (test-present categories only).}
\label{tab:bucket}
\begin{tabular}{lcccc}
\hline
Bucket & BERT & RAG & Hybrid \\
\hline
Zero-shot (0 train) & 0.000 & 0.556 & \textbf{0.619} \\
Rare (1--5)         & 0.000 & 0.277 & \textbf{0.342} \\
Medium (6--20)      & 0.202 & 0.239 & \textbf{0.255} \\
Common ($>$20)      & 0.448 & 0.400 & \textbf{0.457} \\
\hline
\end{tabular}
\end{table}

\subsection{Augmentation Experiments}
We evaluated two data synthesis strategies targeting the label scarcity problem. Table~\ref{tab:cls_aug} reports classifier augmentation results in oracle detection mode (i.e., classification applied to ground-truth positive paragraphs only), isolating classifier performance from detector errors. Table~\ref{tab:det_aug} reports the effect of hard-negative retraining on detector performance and end-to-end hybrid Micro F1.

Augmenting the training set with category descriptions (dict augmentation) slightly reduces oracle Micro F1 (0.651 vs. 0.660 baseline), suggesting that descriptions introduce noise rather than useful signal. Co-occurrence-aware synthetic data yields a marginal improvement of +0.010 in oracle mode, which shrinks to +0.006 in the full pipeline setting (0.483 vs. 0.477) due to the detection bottleneck. Hard-negative detector retraining improves precision substantially (+0.064) but at the cost of 101 additional missed paragraphs (FN: 60 → 161), leaving hybrid Micro F1 unchanged while reducing practical coverage.

\begin{table}[h]
\centering
\caption{Classifier augmentation results (oracle detection mode).}
\label{tab:cls_aug}
\begin{tabular}{lc}
\hline
Classifier Variant & Micro F1 \\
\hline
Baseline                     & 0.660 \\
+ Category descriptions      & 0.651 \\
+ Co-occurrence synthetic    & \textbf{0.670} \\
\hline
\end{tabular}
\end{table}

\begin{table}[h]
\centering
\caption{Hard-negative detector retraining results and effect on
the hybrid pipeline Micro F1.}
\label{tab:det_aug}
\begin{tabular}{lcccc}
\hline
Detector & Pr & Re & F1 & Hybrid \\
\hline
Baseline         & 0.631 & 0.920 & 0.748 & \textbf{0.524} \\
+ Hard negatives & 0.695 & 0.786 & 0.738 & 0.524 \\
\hline
\end{tabular}
\end{table}

\section{Discussion}

\subsection{RAG Strengths, Weaknesses, and the Hybrid Choice}
The bucket analysis in Table~\ref{tab:bucket} reveals a clear complementarity between BERT and RAG. BERT achieves F1 of zero on both zero-shot and rare categories by construction: a discriminatively fine-tuned model has no mechanism for predicting labels it has never encountered during training. To test whether this limitation could be remedied within the supervised paradigm, we augmented the BERT classifier's training set with natural language descriptions of all 135 mitigation categories. Rather than improving zero-shot coverage, this intervention slightly reduced overall oracle Micro F1 from 0.660 to 0.651, with zero-shot category F1 remaining at zero. Category descriptions, when treated as additional training instances, appear to introduce noise without providing the reasoning capacity needed for unseen labels.

RAG addresses this problem at a more fundamental level. By conditioning classification directly on retrieved category definitions at inference time, the pipeline achieves a zero-shot bucket Macro F1 of 0.556 — matching categories that no amount of supervised augmentation could recover. This zero-shot capability is the clearest success of the RAG approach and the primary motivation for the hybrid design.

RAG is not without weaknesses, however. On common categories — those with more than 20 training examples — fine-tuned BERT outperforms RAG (0.448 vs. 0.400). With sufficient labeled data, discriminative fine-tuning learns category-specific linguistic patterns more precisely than retrieval-augmented prompting, which must rely on the LLM's interpretation of category definitions that may not fully capture the nuances of regulatory language. The hybrid pipeline exploits this complementarity: BERT detection provides high recall, ensuring broad candidate coverage, while RAG classification handles the full label space including the long tail. The result is the best performance in every support bucket.

\subsection{Limitations}
Given the structural label scarcity in this domain, we investigated two data synthesis strategies as potential remedies. Results are summarized in Table~\ref{tab:cls_aug} and Table~\ref{tab:det_aug}.

The first strategy augmented the classifier's training set with natural language descriptions of all 135 categories (dict augmentation) and with co-occurrence-aware synthetic paragraphs targeting the 19.4\% of annotated paragraphs that carry multiple mitigation labels. As shown in Table~\ref{tab:cls_aug}, description augmentation slightly reduces oracle Micro F1 from 0.660 to 0.651, suggesting that category descriptions introduce noise rather than useful signal when treated as training instances. Co-occurrence-aware synthesis yields a marginal oracle improvement of +0.010, which shrinks to +0.006 in the full pipeline setting (0.483 vs. 0.477) due to the detection false positive bottleneck: the 406 spurious candidates produced by the BERT detector dilute any classifier-side gain.

The second strategy retrained the BERT detector with 300 synthetically generated hard-negative paragraphs across six types designed to resemble mitigation language without constituting forward-looking obligations. As shown in Table~\ref{tab:det_aug}, this intervention improves detector precision by 0.064 but reduces recall by 0.135, raising the number of missed mitigation paragraphs from 60 to 161. In a compliance monitoring context where missing an obligation is more costly than over-generating candidates, this trade-off is unfavorable. The original detector is retained.

Taken together, these experiments suggest that the primary bottleneck is structural rather than data-driven: 81 of 135 categories are absent from the 2017 test documents, and the detection false positive rate sets a ceiling on downstream classification gain that augmentation alone cannot overcome.

\section{Conclusion}
We presented a hybrid pipeline for automated extraction and classification of environmental mitigation obligations from FERC hydropower licensing documents. The system pairs a BERT-based binary detector with a retrieval-augmented LLM classifier, combining the precision of fine-tuned discriminative models on well-represented categories with the zero-shot reasoning capability of definition-driven prompting on the long tail. Evaluated on the full set of 2017 FERC license documents — a more comprehensive benchmark than previously reported — the hybrid achieves a Micro F1 of 0.524, outperforming both the BERT-only pipeline (0.477) and the RAG-only pipeline (0.416) in every training-support bucket.

A central finding of this work is that the label scarcity inherent in this domain cannot be resolved through supervised augmentation alone. Neither category description augmentation nor co-occurrence-aware synthetic data yielded meaningful gains, and hard-negative detector retraining improved precision at an unacceptable cost to recall. The structural bottleneck — 81 of 135 categories absent from the test set and an irreducible false positive load at the detection stage — defines the limits of the current approach.

Future work will focus on two directions. First, expanding the annotated corpus to cover a larger fraction of the 2,000+ FERC-licensed facilities would reduce label scarcity and bring more categories into both training and evaluation. Second, replacing heuristic hard-negative generation with active learning — using model uncertainty to identify the most informative paragraphs for human annotation — may provide a more principled path to improving detection precision without sacrificing recall.


\bibliographystyle{IEEEBib}
\bibliography{refs}

\end{document}